\documentclass[aps,preprint,preprintnumbers,nofootinbib,showpacs]{revtex4}

\usepackage{amsmath}
\usepackage{amssymb}
\usepackage{mathtools}
\usepackage{dsfont}
\usepackage{color}
\usepackage{graphicx,accents}
\usepackage[normalem]{ulem}

\definecolor{indred}{rgb}{0.8, 0.36, 0.36}
\def\bea{\begin{eqnarray}}
\def\eea{\end{eqnarray}}
\def\sea{\nonumber \\&&}
\def\lla{\left\langle}
\def\rra{\right\rangle}

\def\ssc{\scriptscriptstyle}
\def\lsim{\mathrel{\raise.3ex\hbox{$<$\kern-.75em\lower1ex\hbox{$\sim$}}} }
\def\gsim{\mathrel{\raise.3ex\hbox{$>$\kern-.75em\lower1ex\hbox{$\sim$}}} }
\newcommand{\bra}[1]{\lla#1\right|}
\newcommand{\ket}[1]{\left|#1\rra}
\newcommand{\pdv}[2]{\frac{\partial#1}{\partial#2}}

\makeatletter
\DeclareRobustCommand{\cev}[1]{%
  \mathpalette\do@cev{#1}%
}
\newcommand{\do@cev}[2]{%
  \fix@cev{#1}{+}%
  \reflectbox{$\m@th#1\vec{\reflectbox{$\fix@cev{#1}{-}\m@th#1#2\fix@cev{#1}{+}$}}$}%
  \fix@cev{#1}{-}%
}
\newcommand{\fix@cev}[2]{%
  \ifx#1\displaystyle
    \mkern#23mu
  \else
    \ifx#1\textstyle
      \mkern#23mu
    \else
      \ifx#1\scriptstyle
        \mkern#22mu
      \else
        \mkern#22mu
      \fi
    \fi
  \fi
}

\makeatother
\begin{document}


\title{\boldmath  Quantum Mechanics in Curved Space(time) 
with a Noncommutative Geometric Perspective}

\author{Otto C. W. Kong (otto@phy.ncu.edu.tw)}  

\affiliation{Department of Physics and Center for High Energy and High Field Physics,
National Central University, Chung-li, Taiwan 32054  \\
\vspace*{.2in}}


\begin{abstract}
\vspace*{.2in}
We have previously presented a version of the Weak Equivalence Principle 
for a quantum particle as an exact analog of the classical case, based on the
Heisenberg picture analysis of free particle motion. Here, we take that to
a full formalism of quantum mechanics in a generic curved space(time). Our
basic perspective is to take seriously the noncommutative symplectic geometry 
corresponding to the quantum observable algebra. Particle position coordinate 
transformations and a nontrivial metric assigning an invariant inner product
to vectors, and covectors, are implemented accordingly. That allows an analog 
to the classical picture of the phase space as the cotangent bundle. The 
mass-independent quantum geodesic equations as equations of free particle 
motion under a generic metric as a quantum observable are obtained from an 
invariant Hamiltonian. Hermiticity of momentum observables is to be taken as 
reference frame dependent. Our results have a big contrast to the alternative 
obtained based on the Schr\"odinger wavefunction representation which we 
argue to be less appealing. Hence,  the work points to a very different approach 
to quantum gravity, plausibly with a quantum Einstein equation suggested. 
\end{abstract}


\maketitle

\section{introduction}

Quantum mechanics in curved space(time) is a theory that sits between the 
 well-established theory of basic quantum mechanics and the very challenging 
 theory of quantum gravity. Classical gravity is encoded in the metric, which is firstly 
 given geometrically as what defines an inner product among the (tangent) vectors, 
 and then covectors, independent of their coordinate components. A fundamental 
 notion of a physical vector quantity is that it has an invariant magnitude square, as 
 the inner product with itself. It is firstly through the invariant magnitude square of the 
 physical momentum vector $|\mbox{\boldmath $p$}|^2$, that gravity, or curvature 
 of space(time) affects the dynamics of a particle. The free particle Hamiltonian is 
 $\frac{|\mbox{\boldmath $p$}|^2}{2m}$ which gives motion along geodesics as 
 solutions. In our opinion, the role of the metric is the core of its physical meaning  
 in the particle theory. Our formalism successfully maintains those features and 
 accommodates an exact Weak Equivalence Principle we illustrated in a previous 
 publication \cite{101}.  Apparently, the formalism is incompatible with the
 Schr\"odinger wavefunction representation in general. We contrast our formalism
 against the alternative available \cite{Dw}, based on the latter approach, and discuss
 why that is considered theoretically less desirable, especially as it loses the notion
 of physical vector quantities with invariant magnitude square. 

We have presented the Weak Equivalence Principle for a quantum particle under a constant
gravitational field as an exact quantum version of the classical results, in the language of
covariant Hamiltonian dynamics in the Heisenberg picture \cite{101}. There is no operator 
ordering ambiguity going from the classical to the quantum formulation for that problem 
even in terms of the position and momentum observables of the accelerated frame 
(with Kittel-M\"oller position coordinates). Our Hamiltonian formulation is a consistent 
mathematical description of the classical and quantum geodesic problem as a variational 
problem. The quantum geodesic equations as Heisenberg equations of motion for a free 
particle, in the accelerated frame, are exact operator analogs of the classical ones, with 
nontrivial operator ordering consistent with the following generic form of the vanishing 
of the expression $GE_a$ given, in the same form as the general classical case, by
\bea\label{ge}
GE_a = \frac{d^2{x}^\mu}{ds^2} + \frac{d{x}^\nu}{ds} \Gamma^\mu_{\nu\sigma}  ({x}) \frac{d{x}^\sigma}{ds}  \;.
\eea
Note that the Christoffel symbols are functions of the commuting position observables, 
yet they generally fail to commute with the velocity observables of $\frac{d{x}^\mu}{ds}$.
{Whether the velocity observables commute among themselves is another question.}
The quantum Poisson bracket 
is the commutator divided by ${i\hbar}$. The canonical moment components are $p_\mu$,
and $m \frac{d{x}^\mu}{ds} = p^\mu = g^{\mu\nu}\!(x) \,p_\nu$ can have nontrivial Poisson 
brackets among themselves. The present article presents the general picture of quantum 
mechanics in curved space(time) with any expression for the metric tensor with 
components as arbitrary functions of the position observables. An expression of the 
quantum geodesic equation in terms of Christoffel symbols is, however, not feasible
in the final results. 

We first focus, however, on the `nonrelativistic' case, {\em i.e.} for a metric with an
Euclidean signature, in three dimensions. Basically, we are looking into the problem of
a consistent formulation of quantum mechanics in a curved space. In our opinion, the
problem is of fundamental importance and should be seen as a necessary foundation
for any theory of quantum gravity, but has not received the attention it deserves. 
Apparently, only a 1952 paper from DeWitt\cite{Dw} addresses it seriously, though
it is only explicitly named as a presentation of the `nonrelativistic' dynamics under
a generic position coordinate picture. The work 
has not been discussed often enough in the literature of the subject matter, not to
mention being critically reexamined. DeWitt's starting point is the Schr\"odinger
wavefunction representation, which is the opposite end from our perspectives. The
formalism, in our opinion and as illustrated below, has a few less-than-desirable 
basic features and a quite undesirable result. Most importantly, one cannot have 
the velocity and momentum observables as vector quantities, having a notion of an 
invariant magnitude. In that case, we see the metric as losing most, if not all, of its 
physical meaning in the particle theory. Physics is, of course, about getting 
the `correct' theories, ones that successfully describe Nature as probed through 
observations and experiments in the relevant domains. Yet, within the confine of
the latter, most theorists would see it right to go after `simplicity and beauty', so 
to speak. This is mostly about the conceptual picture and the mathematical form 
of the theories, probably also some of their key results. An unifying feature is an 
important example. It is more desirable to see unifying principles and themes among 
different theories. For the subject matter at hand, it is more desirable to find principles
and features shared by our classical gravitational theory and quantum theory. Those 
should be seen as the more reliable starting point for pursuing the theory of quantum 
gravity. Most would also welcome a quantum theory the metric observable of which 
maintains more of its key features in the classical theory. 

Let us recall the basic perspectives of our approach to the problem and  quantum 
theories in general, emphasizing how that may share common themes and principles 
with the theory of General Relativity (GR). We emphasize a point of view based 
firstly on the observables. They are the physical quantities, period. It is not difficult 
to see mathematical relations among the quantum observables would directly give
the corresponding classical results when the observables are taken as classical ones.
Recovery of the equation of motion for a classical state from the Schr\"odinger equation,
is a less trivial business. The two certainly do not share the same form.  Our second 
key theme is dynamics as given under a Hamiltonian formulation. From the formulation, 
apart from the identification of the physical Hamiltonian among generic Hamiltonian 
functions, the mathematical structure of the system is fully dictated by the symplectic 
geometry of the phase space. So, we have dynamics from geometry, as in GR. The 
phase space is mostly conveniently described in terms of canonical coordinates. For 
a  (spin zero) particle, they are the canonical pairs of position and momentum variables. 
For the real variables in classical mechanics, we have the basic Poisson bracket relations
\bea\label{pb}
\{ x^i, x^j \} =0 = \{p_i, p_j \} \;,
\qquad
\{ x^i, p_j \} =\delta^i_j \;.
\eea
The expressions are independent of even the existence, not to say the form, of a metric 
tensor. Note that the generic Poisson bracket relations can be verified through an 
implementation of a transformation of the position coordinates from the simple exact 
Euclidean ones. The particle phase space can be seen as the cotangent bundle of the 
configuration/position space, hence having $p_i$ as components of the momentum 
vector as firstly a covector, though $x^i$ as generic position coordinates in a curved 
space cannot be taken as components of any vector. The cotangent bundle of 
a manifold, with or without a Riemannian structure, has a natural symplectic geometry. 
{The symplectic structure gives} curves of Hamiltonian flow for a generic Hamiltonian 
function $H_{\!s'}$ as integral curves of the Hamiltonian vector field $\{ \cdot, H_{\!s'} \} = \frac{d}{ds'}$ 
where the mathematical nature of the flow parameter $s'$ is dictated by $H_{\!s'}$. 
For classical mechanics within the formulation, the only proper way to look at the 
Newtonian time is that it is such a parameter for physical Hamiltonian. The 
Hamiltonian flow is then to be interpreted as the physical time evolution. While the
position variables as part of the phase space coordinates are part of the kinematic
set-up of the theory, with their validity subjected to the validity of the theory, time is 
completely about the dynamics of specific systems. The picture of Newtonian space-time 
as a coset space of the Galileo group, and the exact idea of the latter as the relativity 
symmetry behind Newtonian mechanics is not justified and indeed incorrect \cite{095}. 
Though the formulation as given through Newton's Laws requires a model for space
and time as the starting point, Galilean time translation can only give the correct
picture for a particle when it is free, but not when it has nontrivial dynamics. From 
the point of view of Hamiltonian mechanics, relativity symmetry should be about 
reference frame transformations for the phase space. 
 
Here, we are talking about the metric tensor in the usual sense, {\em i.e.} as one firstly 
for the configuration space. In the case of a classical particle, 
 the configuration space is our physical space or rather our classical model of it. 
We want to emphasize the logic that it is from 
our successful theory of particle dynamics that we can retrieve from it a successful 
model of our physical space. That is to be seen as the totality of all possible 
positions of a particle. Any otherwise notion of (empty) space in itself is not physical. 
For quantum mechanics, we see no justification to maintain the idea that the classical 
model of our physical space is the proper model to look at Nature as described by 
the theory. The phase space for a quantum particle certainly does not have the 
three-dimensional real manifold as a configuration subspace. We have presented
a picture of the quantum phase space as a noncommutative symplectic geometry with 
the quantum position and momentum observables as canonical coordinates \cite{078}.
The Poisson bracket is the standard one, the commutator divided by $i\hbar$. From
the discussion above, one would conclude that the picture should remain valid with 
the introduction of a nontrivial metric with components as functions of the position 
observables. Hence, we should have quantum position and momentum observables 
satisfying the canonical Poisson bracket conditions of Eq.(\ref{pb}). In the general 
mathematical theory of noncommutative geometry \cite{C,Ma,c,Dv}, Riemannian 
structure has been analyzed \cite{c}. While the language in the latter literature hardly 
fits our purpose, and it is not trivial to see if there could be such a picture (of the 
so-called spectral geometry) for the position subspace of our quantum phase space,
some common features to the theory of quantum mechanics we are seeking here is
worth attention. Connes \cite{c} has geodesic flows as a one-parameter group of
automorphism on the (observable) algebra, with a generator as essentially the magnitude 
square of a cotangent vector. Such a group of automorphism is exactly one of unitary 
transformation in the Schr\"odinger picture. They are Hamiltonian, with a generator as 
the free particle Hamiltonian proportional to the magnitude of the canonical momentum 
vector, an exact analog of the classical, commutative, case of $H= \frac{1}{2m} p_i g^{ij} p_j$. 
As the readers will see, all the features discussed are what we want, and have 
successfully retained in our formulation presented below.

We introduce next the notion of Pauli metric operator (PMO), denoted by $g_{\!\ssc P}$.
In the classical geometric picture of the quantum phase space as the (projective) Hilbert 
space, it is a K\"ahler manifold with the metric structure tied to the symplectic structure. 
The feature cannot be maintained in a noncommutative geometric picture, as the `coordinate 
transformation' between the two pictures does not respect the complex structure \cite{078}. 
The two pictures are essentially the Schr\"odinger and the Heisenberg pictures. Following
a study by Dirac \cite{D}, basically trying to put a notion of Minkowski metric onto a vector 
space of quantum states, Pauli introduced the metric operator that defines the inner product 
as $\bra{\cdot}  g_{\!\ssc P} \ket{\cdot}$ \cite{P}, that is also the key feature
behind the more recent notion of pseudo-Hermitian quantum mechanics \cite{pH}. When 
the PMO is trivial, we have the usual Hilbert space for quantum mechanics, in flat space. 
What is important to note is that the proper notion of operator Hermiticity for an inner 
product space is to be defined relative to the inner product. 
Explicitly, with the notation
$\prescript{}{g_{\ssc p}}{\!\!\lla  \cdot |  \cdot \rra}\equiv \bra{\cdot}  g_{\!\ssc P} \ket{\cdot}$,
the Hermitian conjugate of operator $A$ relative to the inner product is given by the
operator ${A}^{\dag^{g_{\ssc p}}}$ satisfying the equation. 
\bea 
\prescript{}{g_{\ssc p}}{\!\!\lla  \cdot | {A}^{\dag^{g_{\ssc p}}} \cdot \rra}
= \prescript{}{{g_{\ssc p}}}{\!\!\lla  {A}\cdot | \cdot \rra} \;.
\eea
In terms of the naive Hermitian conjugate for the case with a trivial metric, which
we denote by the usual $A^{\dag}$, we have 
\bea\label{gH}
{A}^{\dag^{g_{\ssc p}}}= g_{\!\ssc P}^{-1} {A}^\dag {g_{\!\ssc P}} \;.
\eea  
Note that while we can have many different inner products introduced for a vector space,
only one can be considered the physical inner product, as there is one physical metric.
In a theory of `relativistic' quantum mechanics \cite{087}, we have illustrated an explicit 
notion of a Minkowski metric of the vector space of state having a Minkowski PMO $\eta$ 
that is directly connected to a naive notion of a Minkowski metric tensor in the 
noncommutative geometric picture. We have position and momentum observables 
$X^\mu$ and $P^\mu$ being $\eta$-Hermitian satisfying 
$X^\mu = \eta X_\mu \eta^{-1} = \eta^{\mu\nu} X_\nu$ and
$P^\mu = \eta P_\mu \eta^{-1} = \eta^{\mu\nu} P_\nu$, with $\eta_{\mu\nu}=\{-1,1,1,1\}$.
In the naive Schr\"odinger representation which Pauli mostly worked with, for the 
`nonrelativistic' case, we have $g_{\!\ssc P} (x)$ as $ \sqrt{|\det{g_{ij}(x))}|}$, from the 
invariant volume element \cite{PB}. The key point is that the notion of Hermiticity may 
generally be metric-dependent, hence dependent on the choice of coordinates. In the spirit 
of GR, the choice of frame of reference with the coordinates as part of it is like a gauge choice. 
The coordinate observables are then gauge-dependent quantities. We are using the term 
observables in the more general sense as in the literature of mathematical physics, without 
requiring them to be Hermitian in any sense. We use the term physical observables for 
observables as described in basic quantum mechanics textbooks, hence Hermitian. There is 
nothing so unphysical about a non-Hermitian element of the observable algebra though. A 
simple product of two Hermitian observables which do not commute would not be Hermitian.
As to be presented below, we have various notions of Hermiticity on the same quantum 
observable algebra, each defined by a PMO, referred to as $g_{\!\ssc P}$-Hermiticity,
relevant in a different coordinate picture of the related noncommutative geometry. Now, 
if what can be considered a physical (gauge-dependent) observable depends on the gauge 
choice, it may not be considered unreasonable. 

Another important feature of the noncommutative geometric picture we want to mention
here is the notion of quantum reference frame transformations \cite{qf,093}. We are 
concerned here with how to think about the dynamical theory from different systems 
of coordinates or frames of reference. The focus is on a generic transformation of the
quantum position coordinates as $x^i \to  x'^a = x'^a (x)$, as the exact parallel of what 
we have in classical physics. The position observables are taken as functions of the old 
ones in the usual classical sense. Then all the position observables old and new commute 
with one another. When we talked about going to an accelerated frame in Ref.\cite{101}, 
for example, the relative acceleration between the old and new frames is essentially 
taken as a classical quantity. It was assumed at least to commute with the position 
observables. Such coordinate transformations are classical in nature, though they have
quite nontrivial implications on momentum observables as quantum quantities.
A full consideration of the free-falling frame of a quantum particle 
needs to consider the gravitational acceleration as a quantum quantity. 
{Solving the problem may be still quite a challenge to be taken.}  In a quantum spatial 
translation position eigenstates are not necessarily taken to position eigenstates,
as well illustrated in Refs.\cite{qf,093}. In particular, we have presented a way to look at 
a simple quantum spatial translation, as needed for the description of the position of 
a quantum particle relative to another quantum particle in a generic state. In terms of the 
full information about the quantum position of 
{the particles seen in the original frame,} 
an explicit description of the quantum quantities about the exact `amounts' the position 
observables of the other particle differs between the old and new frames have been
presented \cite{093}. Heisenberg uncertainties in the position observables and related
features of entanglement changes are fully encoded in those exact `amounts'. The 
important relevancy of quantum reference frame transformations to a theory of quantum 
gravity has been addressed by Hardy\cite{H}. After all, the Relativity Principle has a 
fundamental role in GR. Nature does not use any frame of reference and is not expected
to prefer anyone over another. Any practically defined frame of reference has a physical 
nature and hence relies on the physical properties of matter that are quantum in nature. 
Penrose has emphasized a supposed incompatibility between quantum mechanics and 
the Relativity Principle \cite{Pn}. Essentially, it is the observation that no reference frame 
transformation can take a position eigenstate to one that is a nontrivial linear combination 
of such eigenstates. However, that is true only when we limit ourselves to transformations 
that are classical in nature. Otherwise, it can easily be done with a simple quantum 
translation. A particle in a position eigenstate would not be in an eigenstate of its position
as observed from another, observing, particle that in itself is not in an eigenstate. 
A quantum translation from the original frame to the frame of the observing particle would 
change the nature of the first, observed, particle as a position eigenstate or not in a way 
dependent on the corresponding properties of the second particle. Such reference frame 
transformations cannot be pictured in any classical/commutative geometry. But they are 
natural in our noncommutative geometric picture \cite{078}.

In the next section, we discuss the various key issues from the perspective of 
noncommutative symplectic geometry for the phase space corresponding to the 
quantum observable algebra. DeWitt's formalism is also sketched to illustrate what 
we see as limitations and unpleasant features of it. Analyzing all that sets a clear
platform for the appreciation of the key features of our formalism, {which is} 
presented in Sec.3, focusing on the case of Euclidean metric signature. The 
corresponding picture for the case with Minkowski metric signature incorporating 
our results of Ref.\cite{101} is addressed in the section after. Concluding remarks 
are presented in the last section.

 \section{Noncommutative Symplectic Geometry, Metric, Hermiticity, and Coordinate Transformations}
The mathematical starting point is the noncommutative quantum observables algebra.
For mathematicians, any abstract $C^*$-algebra has a corresponding geometry
admitting a Riemannian structure \cite{C}. The algebra may be seen as the algebra 
of some kind of functions on the geometric space, which would be a noncommutative
one for a noncommutative algebra. An algebra of physical interest is an algebra of 
observables, which are generally considered to be $C^*$-algebras \cite{S}. Those
would be algebras of `observables' that include complex linear combinations 
of all physical observables. For a particle without spin, with an extension beyond
beyond the formal bounded limit, it is naturally to be considered 
as formal complex functions of the position and momentum observables. 
So, while the mathematicians have not quite studied symplectic geometry in the
noncommutative case and hardly considered coordinates, a picture of symplectic 
geometry with noncommutative canonical coordinate observables satisfying the 
Poisson bracket conditions of Eq.(\ref{pb}) naturally comes to our attention. We
believe the picture, with the standard quantum Poisson bracket, is what goes 
along with the true spirit of thinking about quantum mechanics first introduced by
Heisenberg and Dirac. Every $^*$-algebra necessarily has a notion of Hermitian 
conjugation, or called an involution, and can be seen as an algebra of operators on 
a Hilbert space. But as discussed above, we want to look at the physics picture 
allowing the use of different systems of coordinates related to one another through 
generic position coordinate transformations for the quantum position observables of 
the classical kind, $x^i \to  x'^a = x'^a (x)$ where all position observables commute
among themselves. With each set of position coordinates, we have a metric tensor
with components {being functions of them.} Pauli's study tells us that in each 
case, the relevant notion of Hermiticity is the specific $g_{\!\ssc P}$-Hermiticity. 
We may not assume Pauli's explicit expression of $g_{\!\ssc P} (x)$ as
$\sqrt{|\det{g_{ij}(x))}|}$ to work beyond a Schr\"odinger wavefunction 
representation, at least we have to admit the possible inequality between 
$g_{\!\ssc P} (x)$ and $g_{\!\ssc P} (x')$ and hence different notions of the
Hermiticity. The explicit lack of invariance of $\sqrt{|\det{g_{ij}(x))}|}$ is an indication of that. 
{We can keep the  Schr\"odinger picture of the observables as operators on 
a vector space of states at least locally.} Yet, the naive inner product and the related notion 
of Hermiticity may not necessarily be physically relevant. The naive inner product is 
one given by an Euclidean metric among the position coordinate observables with 
the PMO given by the identity operator. Under a different set of coordinates, the 
metric expression changes. A different PMO is expected giving a different inner
product and different notion of $g_{\!\ssc P}$-Hermiticity. We have to look
explicitly into how the quantum observables transform and hence have to deal with 
different notions of Hermiticity at the same time. In our approach to the problem, 
we first take all observables as algebraic functions of the position coordinate 
observables to be Hermitian, indeed under all $g_{\!\ssc P}$, though the only 
Hermiticity to be concerned about would be the $g_{\!\ssc P} (x)$-Hermiticity 
when the canonical position coordinate observables are taken to be $x^i$.  The
PMO $g_{\!\ssc P}(x)$ has generally taken to be different in a different frame of 
reference, or a different choice of such system of coordinates, though we do not 
have to commit to a particular form of each such  PMO. The 
$g_{\!\ssc P} (x)$-Hermiticity of the canonical momentum coordinate observables
$p_i$ and that of $p^i$ we leave open, to be addressed till after the major 
mathematical analysis. 

With a set of quantum canonical coordinates observables, the key question is how 
we can have a picture of quantum mechanics with a physical Hamiltonian of a free
particle that would be an invariant quantity under any position coordinate 
transformation, and preferably maintains its classical relation with the canonical
momentum. A look at DeWitt's paper \cite{Dw} tells us that it is nontrivial. We 
present here DeWitt's formalism and what we see as its shortcomings to illustrate 
the key issues. That helps to set the stage for the presentation of the theory from 
our new approach.

Sticking to Schr\"odinger wavefunction representation largely thinking in terms of 
writing down the theory in a generic coordinate system as from position coordinate 
transformation $x^i(x')$ from Euclidean $x'^a$, DeWitt took as the invariant free 
particle Hamiltonian $\frac{-\hbar^2}{2m} \Delta \equiv \frac{-\hbar^2}{2m} \nabla_{\!i} g^{ij} \nabla_{\!j}$, 
with the invariant Laplacian $ \Delta$. Yet, he realized that one cannot take $-i\hbar \nabla_{\!i} $ 
as the canonical momentum. Firstly, the coordinate covariant derivatives do not commute 
among themselves. While the Laplacian has the form $\nabla_{\!i} g^{ij} \nabla_{\!j}$,
thinking about it as something like the magnitude square of a cotangent vector does
not make good sense. $-i\hbar \partial_i$ certainly does work either, and  
$\partial_i g^{ij} \partial_j$ is not even invariant. Through implementing the canonical
condition effectively as Heisenberg commutation relations onto the properly normalized 
position eigenstate wavefunctions, DeWitt arrived at the canonical momentum
$p_i^{\ssc D}$ as given by
\bea
p_i^{\ssc D} = -i\hbar \left( \partial_i + \frac{1}{2} \Gamma^k_{ki} \right) .
\eea
The set of $x^i$ and $p_i^{\ssc D}$ not only satisfy the condition of Eq.(\ref{pb}),
when a generic transformation $x^i \to x'^a(x), p_i^{\ssc D} \to  p'^{\ssc D}_a$ 
with general non-Euclidean $x'^a$ is taken, those conditions hold for new 
coordinates. The actual transformation rule for DeWitt's momentum is given by
\bea\label{Dt}
p'^{\ssc D}_a = \frac{1}{2}    \left(  \pdv{x^i}{x'^a} p_i^{\ssc D} + p_i^{\ssc D} \pdv{x^i}{x'^a}  \right) .
\eea
He took it as naturally the necessary symmetrized form of the classical version, to 
maintain Hermiticity, {\em i.e.} Hermiticity of  $x^i$ and $p_i^{\ssc D}$ lead to 
Hermiticity of $x'^a$ and $p'^{\ssc D}_a$. The first problem is $\frac{-\hbar^2}{2m} \Delta$ 
generally differs from $\frac{1}{2m} p_i^{\ssc D} g^{ij} p_j^{\ssc D}$. 
{The free particle 
Hamiltonian expressed in terms of that `magnitude square of the momentum vector'} 
has an extra term which is 
$\frac{\hbar^2}{2m}$ times a somewhat complicated expression involving 
{the inverse 
metric tensor, Christoffel symbols and the coordinate derivatives of them,} but not 
 $p_i^{\ssc D}$. DeWitt called it a quantum potential. The idea that quantum mechanics
gives such an extra potential to a free particle is very unpleasant. What would 
correspond to the quantum geodesic equation looks even more complicated, of course, 
with mass-dependent terms. Pauli suggested the same canonical momentum under
a different consideration, one of $g_{\!\ssc P} (x)$-Hermiticity as discussion in the
introduction section. In the expression given by Pauli   \cite{PB}, we have
 \bea
 p_i^{\ssc D} = - \frac{1}{\sqrt{g_{\!\ssc P} (x)}} i\hbar \partial_i  \sqrt{g_{\!\ssc P} (x)} \;.
 \eea
The $g_{\!\ssc P} (x)$-Hermiticity is obvious when applying Eq.(\ref{gH}). In his
consideration of the properly normalized position eigenstate wavefunctions, DeWitt 
had effectively put in the notion of PMO $g_{\!\ssc P} (x)$ as $ \sqrt{|\det{g_{ij}(x))}|}$.
Now, $p_i^{\ssc D}$ is $g_{\!\ssc P} (x)$-Hermitian while $p'^{\ssc D}_a$ is
 $g_{\!\ssc P} (x')$-Hermitian, $g_{\!\ssc P} (x) \ne g_{\!\ssc P} (x')$. With the different 
 notions of Hermiticity for the canonical momentum observables relevant under the 
 different frames of reference, the consideration of adopting a symmetrized form for 
 the transformation rule sounds meaningless. There seems to be an inconsistency 
 obtained in putting all that together. The transformation rule of Eq.(\ref{Dt}) 
 suggests $p'^{\ssc D}_a $ is also $g_{\!\ssc P} (x)$-Hermitian, which is 
 incompatible with Eq.(\ref{gH}).
 
As discussed above, the classical picture is that the canonical momentum
is a cotangent vector, hence transforms as one. But $p_i^{\ssc D} g^{ij} p_j^{\ssc D}$ 
is not an invariant; it does not equal to  $p'^{\ssc D}_a g^{ab} p'^{\ssc D}_b$. Then 
there cannot be any notion of $p_i^{\ssc D}$ as a covector and the inverse metric $g^{ij}$
{as defining an inner product on the momentum (co)vector.}
As $x^i$ is not a vector, essentially the 
only notion of the idea of a metric as in classical geometry involved in the Hamiltonian
formulation of the dynamics is in that notion of an invariant magnitude square of the 
canonical momentum as a covector. Giving that up, it is questionable if there is still 
any useful notion of metric for the theory. A related question is about $p^{\ssc D^i}$. 
Requiring both $p^{\ssc D^i}$ and $p_i^{\ssc D}$ to be Hermitian, under whatever
notion of Hermiticity, one has again to go with a symmetrized form of their relation, 
such as $p^{\ssc D^i} = \frac{1}{2}  \left(  g^{ij} p_j^{\ssc D} + p_j^{\ssc D} g^{ji} \right)$.
And DeWitt's free particle Hamiltonian does gives $m \frac{dx^i}{dt}$ equals to the
expression. But then $p_i^{\ssc D} g^{ij} p_j^{\ssc D}$, $p^{\ssc D^i} g_{ij} p^{\ssc D^j}$, 
$p^{\ssc D^i} p_i^{\ssc D}$, $p_i^{\ssc D} p^{\ssc D^i}$, and 
$\frac{1}{2}( p^{\ssc D^i} p_i^{\ssc D}$, $p_i^{\ssc D} p^{\ssc D^i} )$ would all be 
different quantities, none of them being invariant. 
The velocity $\frac{dx^i}{dt}$ we want to be considered a vector, again as a basic part
of the notion of a metric seen by the quantum particle. So, we can see that the usual 
notion of vector and covector related by the metric tensor and its inverse with invariant 
magnitude seems to be incompatible with having all such quantities to be Hermitian, 
whatever notion of Hermiticity applied. Keeping components of a vector observable
and those of its covector to be both Hermitian, in particular, is not compatible with the
notion of its having an invariant magnitude square given through the metric tensor.

DeWitt had probably given the best possible formalism within the Schr\"odinger wavefunction 
representation. We have stated the shortcomings of the formalism. It is then interesting to 
see if a better picture can be obtained from a Heisenberg picture consideration with our 
noncommutative geometric perspective. Note that while one would always have a matching 
Schr\"odinger picture with states to be described as vectors in a vector space, that is not
necessarily the same as having a Schr\"odinger wavefunction representation. 
{One can reasonably question if the  wavefunction representation can work} 
for a curved space or in an arbitrary coordinate
picture of the position observables. We have already expressed our criticism of the traditional 
view of seeing the classical space coordinated by the real variables as arguments for the 
Schr\"odinger wavefunction as the valid model for space on which a quantum particle moves.
After all, the quantum position coordinate variables are elements of the noncommutative
observable algebra. They are, in Dirac's words,  q-number quantities that take q-number
values \cite{D27}. The q-numbers would be elements of a noncommutative algebra. Going 
beyond that traditional view we see little problem in giving up the Schr\"odinger wavefunction 
representation.

\section{Our Approach to Quantum Mechanics in Curved Space}
Physical quantities as dynamical variables in a quantum theory are elements of 
a noncommutative algebra, with the extra Poisson bracket structure, hence having
a noncommutative symplectic geometric picture. Looking into the literature on
introducing calculus into noncommutative geometry, we note a few basic features.
First of all, from the algebraic point of view, a derivation (something that acts like
a differential operator) is to be given through a commutator action. Good examples
for our case are the Hamiltonian vector fields $\{\cdot,H'\}$ of an observable $H'$.
In particular, we have
\bea\label{hvf}
\partial_i \equiv \pdv{}{x^i} = \{\cdot,p_i \} \;,
\qquad
 \pdv{}{p_i} = -  \{\cdot,x^i \} \;.
\eea
That was how we introduced the coordinate derivatives in Refs.\cite{078,081} and 
also what we use for any derivative with respect to any position observables here.
The actions of such derivations on functions of the observables go formally in the same 
way as for the classical observables, though a function may involve nontrivial variable 
ordering which is to be respected in taking the derivation, as the commutator action 
naturally does. The next to note is that to the extent that coordinates are used, see 
for examples Refs.\cite{Ma,WZ}, with some notion of contracting vector and covector 
indices, no symmetrized form is ever used. Say one has the exterior differentiation 
operator {\boldmath $d$} expanded as {\boldmath $dy^n \pdv{}{y^n}$} for $\{ y^n \}$
as a set of noncommutative coordinates. That suggests for the vector $V^i$ and
covector $W_i$ on the position space as part of the noncommutative symplectic
geometry, we have the scalar quantity $V^i W_i$ with the components transform
as
 \bea\label{tv}
 {V}'^a = {V}^i \pdv{{x}'^a}{{x}^i}\;,
 \qquad
  {W}'_a = \pdv{{x}^i}{{x}'^a} {W}_i  \;,
\eea
maintaining ${V}'^a  {W}'_a =  {V}^i  {W}_i$. The expression 
${W}_i  {V}^i = {V}^i  {W}_i - [{V}^i, {W}_i]$ is then something different so long as the 
observables involved do not commute; and we do not have that equals to ${W}'^a  {V}'_a$
in general. We put the expression for the exterior differential operator above in bold 
type for a reason. As explicitly noted in Ref.\cite{WZ}, 
{ in the naive expansion of {\boldmath $d$} in terms of {\boldmath $\partial_n \equiv \pdv{}{y^n}$}, 
\[ \mbox{ 
{\boldmath $dy^n \partial_n$}$ ff'=$ {\boldmath $d$} $ff'=$({\boldmath $d$}$f)  f' +
  f${\boldmath $d$}$f'=(${\boldmath $dy^n \partial_n$}$f) f' + f${\boldmath $dy^n \partial_n$}$f'$
}\]
generally does not give {\boldmath $\partial_n$}$ff'$ as 
  ({\boldmath $\partial_n$}$f)  f' + f${\boldmath $\partial_n$}$f'$
unless the commutator $[ \mbox{\boldmath $dy^n$}, f ]$ vanishes.} Without knowing
the basic commutation relations between $y^m$ and {\boldmath $dy^n$}, there is no
way to have an explicit expression for the action of {\boldmath $\partial_n$}. In fact,
unless the commutator $[f ,\mbox{\boldmath $d^ny$}]$ can be written as   
{\boldmath $d^ny$}$f''$ for some function $f''(y)$, one cannot even write   
{\boldmath $\partial_n$} as an explicit operator on any product function $ff'$. But we 
do not need anything of the kind, only the derivations defined through the Hamiltonian
vector fields of Eq.(\ref{hvf}). The formulation of calculus then is more in line with the
one from matrix geometry \cite{Ma}, indeed explicitly given in
Dubois-Violette\cite{Dv}. One forms are generally well-defined. Yet, so far as
this article is concerned, we do not even have to think about them. We have only a
differentiation as differentiation  with respect to a real parameter, one of Hamiltonian 
flow like the Newtonian time. For example, $\frac{d{x}^{i}}{dt}$ is simply $\frac{d}{dt} x^i$. 
 
With Eq.(\ref{tv}), we have, under any notion of Hermitian conjugation
\bea
 V'^{a\dag} = \left( \pdv{{x}'^a}{{x}^i}  \right)^\dag  {V}^{i\dag} \;,
 \qquad
   {W}'^\dag_a = {W}_i^\dag \left( \pdv{{x}^i}{{x}'^a} \right)^\dag  \;.
\eea
The important point to note is that a vector or covector has different transformation
properties with its Hermitian conjugation. Then even if one has components all of
which are Hermitian, we would have to keep track of that. Explicitly, consider $A^i$ 
as $g_{\!\ssc P} (x^i)$-Hermitian. While $\pdv{{x}'^a}{{x}^i}$ shares the same property, 
${A}'^a = {A}^i \pdv{{x}'^a}{{x}^i}$ would not. And whether the components are 
Hermitian or not, the equation above represents different transformation properties 
compared to Eq.(\ref{tv}), and hence different notions of vectors and covectors. We
introduce a new notion for the indices to keep track of the different transformation
properties, writing generally 
\bea
 V'^{\ssc A} = \pdv{{x}'^{\ssc A}}{{x}^{\ssc I}}    {V}^{\ssc I} \;,
 \qquad
   {W}'_{\!\ssc A}  = {W}_{\!\ssc I}  \pdv{{x}^{\ssc I}}{{x}'^{\ssc A}} \;,
\eea
bearing in mind ${V}^{\ssc I}$ is essentially ${V}^{i\dag}$, and  ${W}_{\!\ssc I}$
is essentially ${W}^{\dag}_i$, and drop any explicit notation for any notion of
Hermitian conjugation. Of course, each $x^{\ssc I}$ is no different from $x^i$.
Note that with the capital indices, admissible index contraction gives an invariant
${W}_{\!\ssc I} {V}^{\ssc I}= {W}_{\!\ssc A} {V}^{\ssc A}$ but not $ {V}^{\ssc I} {W}_{\!\ssc I}$.

A word of caution is necessary here against a more general picture of tensors and the 
vector index contraction. As $V^i U W_i \ne V^i W_i U$ or $U V^i W_i$ in general, one 
cannot think about the former as an invariant scalar quantity even if $U$ is one. Similarly, 
naive notion of tensors as tensor products of vectors and covectors may not work.
However, for expressions involving only the position observables but not their time 
derivatives and the momentum observables, all tensor notions are still valid as there is 
no nontrivial commutation relations to worry about. So, the quantum analog of all the 
geometric tensors of a classical Riemannian manifold maintains their transformation 
properties under the quantum analog. Besides, quantities such as $V^{\!\ssc A} W^a$
and $V'_a  {W}'_{\!\ssc A} = g_{a{\ssc B}}  V^{\!\ssc B} W^b g_{b{\ssc A}}$ are still
proper tensors. For example, a naive consideration would give the quantum Einstein 
equation, for the quantum observables, simply as 
$G_{\!a{\!\ssc A}} + \Lambda g_{a{\ssc\! A}} = \kappa T_{\!a{\ssc\! A}}$.
Again, components of the geometric tensors as commuting products of functions
of position observables are always Hermitian. Hence, those for $G_{\!a{\!\ssc A}}$
are exactly the same as the corresponding $G_{ab}= R_{ab} - \frac{1}{2} R \, g_{ab}$.
Provided that a proper expression for the quantum energy-momentum tensor 
$T_{\!a{\ssc\! A}}$ could be given in various settings, the equation could be the
master equation of a theory of quantum gravity.

With the discussion above and the enriched notation, we are ready to present the 
dynamical theory. In terms of the canonical momentum $p_i$ and its 
$g_{\!\ssc P}(x)$-Hermitian conjugate $p_{\!\ssc I}$, we take the free particle 
Hamiltonian as
\bea
H_t = \frac{1}{2m} p_{\!\ssc I} g^{{\ssc I}j} p_j \;.
\eea
That is an invariant quantity with the Hermitian inverse metric tensor written as 
$g^{{\ssc I}j}$ following our transformation and index contraction rules.
Taking $p^j = p_{\!\ssc I}g^{{\ssc I}j}$ and  $p^{\ssc I} = g^{{\ssc I}j} p_j$, and then
$p_{\!\ssc I} = p^j g_{j{\ssc I}}$ and $p_j = g_{j{\ssc I}} p^{\ssc I} $, we 
\[
p^j p_j = p_{\!\ssc I} g^{{\ssc I}j} p_j =  p_{\!\ssc I} p^{\ssc I} = p^j g_{j{\ssc I}} p^{\ssc I} 
\]
as the invariant magnitude square of the momentum (co)vector. Irrespective of the 
Hermiticity properties of momentum components, it is always Hermitian under any of the 
notions of $g_{\!\ssc P}$-Hermiticity, so is $H_t$. Following 
$\frac{d}{dt} = \{ \cdot, H_t\} \equiv \frac{1}{i\hbar} [ \cdot, H_t]$, we have
\bea\label{dx}
\frac{d}{dt} x^{(i)}= \{ x^{(i)}, H_t\} = \frac{1}{2m} \left( g^{{\ssc I}j} p_j + p_{\!\ssc J} g^{{\ssc J}i} \right)
  = \frac{1}{2m} ( p^{\ssc I} +  p^i ) \;,
\eea
where we have put the bracketed index to emphasize that it is not a vector index and
hence no distinction between $x^{(i)}$ and $x^{(\ssc I)}$. Note that $p^{\ssc I}$ is,
however, the $g_{\!\ssc P} (x)$-Hermitian conjugate of $p^i$, which we allow to be
different at this point to keep the analysis more generally valid. To retrieve a notion 
of the velocity vector, we then introduce
\bea\label{dp}
\frac{dx^i}{dt} = \frac{p^i}{m} \;,
\qquad
\frac{dx^{\ssc I}}{dt} = \frac{p^{\ssc I}}{m} \;, 
\eea
and mutual  $g_{\!\ssc P} (x)$-Hermitian conjugates, giving each
$g_{\!\ssc P} (x)$-Hermitian $\frac{d}{dt} x^{(i)}$ as 
$\frac{1}{2} \!\left( \frac{dx^i}{dt} + \frac{dx^{\ssc I}}{dt} \right)$. We can also have
the definitely $g_{\!\ssc P} (x)$-Hermitian $p^{(i)} = \frac{1}{2}  ( p^{\ssc I} +  p^i )$.
Next, we have
\bea &&
\frac{dp_i}{dt} = \{ p_i, H_t\} = - p_{\!\ssc J} \frac{ \partial_i g^{{\ssc J}k}}{2m} p_k \;,
\sea
\frac{dp_{\!\ssc I}}{dt} = \{ p_{\ssc I}, H_t\} = - p_{\!\ssc J}  \frac{ \partial_{\ssc I} g^{{\ssc J}k}}{2m} p_k\;.
\eea
{We have adopted an economy of notations that has to be read carefully. The two
velocity vectors in Eq.(\ref{dp}) are generally to be taken as Hermitian conjugates.  Even
their corresponding components are not here assumed to be equal.  
{Without the Hermiticity of the above expressions, $\frac{dx^i}{dt}$, or 
$\frac{dx^{\ssc I}}{dt}$ are not to be taken as time derivatives in the usual sense.}
Similarly, we are not committed to an equality of $p_i$ and $p_{\ssc I}$ components.}
(Note that straightly speaking, one should write $\partial_{\!\ssc I} g^{{\ssc J}k}$ in 
the form $g^{{\ssc J}k} \cev\partial_{\!\ssc I}$ to show its transformation property.) 
Hermitian conjugation gives also 
$\{ x^I, p_{\ssc J}\} \equiv \{ x^i, p_{\ssc J}\}^{\dag\dag}= i\hbar \delta^i_{\ssc J}$.
Note that the above results are mostly obtained directly from applying the quantum
Poisson bracket or its Hamiltonian vector fields as derivations. Yet, the `velocity' is only 
given from the Hamiltonian evolution as in Eq.(\ref{dx}). When $p^i$  is not Hermitian,
the $\frac{d}{dt} x^{(i)}$ is not a vector. Hence we are free to introduce the velocity vector
as in   Eq.(\ref{dp}). Eventually, we take having individual $p^i$ components as 
$g_{\!\ssc P} (x)$-Hermitian, hence equals to the corresponding $p^{\ssc I}$, as the right 
choice that can be made consistently. That could have equality of $\frac{d}{dt} x^{(i)}$, 
$\frac{dx^i}{dt}$ and $\frac{dx^{\ssc I}}{dt}$ component-wise, and Eq.(\ref{dp}) is just
Eq.(\ref{dx}). 
{We would not have then equality of the individual $p_i$ and $p_{\ssc I}$. 
Both of them still satisfy the condition of being the canonical conjugate variable of $x^i$ though.}

Putting together the above, one can get to the following equations, in a form closest to the 
classical geodesic equations, for comparison:
  \bea&&              
      \frac{d^2{x}^{i}}{dt^2} 
      = \frac{d{x}^h}{dt}    \frac{\partial_{\ssc J} g_{h{\ssc K}} }{2}  \frac{d{x}^{\ssc K}}{dt} g^{{\ssc J} i}
                        - \frac{d{x}^h}{dt}  \frac{\partial_{\ssc K} g_{h{\ssc J}}}{2}  g^{{\ssc J} i} \frac{d{x}^{\ssc K}}{dt}  
                   - \frac{d{x}^k}{dt}  g_{kM} \frac{d{x}^h}{dt}  g^{Ml } \frac{\partial_h g_{l{\ssc J}}}{2} g^{{\ssc J} i} \;,
  \sea\label{geo}
     \frac{d^2{x}^{\ssc I}}{dt^2}  
     = g^{{\ssc I} j} \frac{d{x}^k}{dt}    \frac{\partial_j g_{k{\ssc H}} }{2}  \frac{d{x}^{\ssc H}}{dt} 
                       -   \frac{d{x}^k}{dt} g^{{\ssc I}j}  \frac{\partial_k g_{j{\ssc H}}}{2}  \frac{d{x}^{\ssc H}}{dt} 
           - g^{{\ssc I} j}   \frac{\partial_{\ssc H} g_{j{\ssc L}}}{2}  g^{{\ssc L}m} \frac{d{x}^{\ssc H}}{dt} g_{m{\ssc K}} \frac{d{x}^{\ssc K}}{dt}  \;.
 \eea 
Again, we have the $g_{\!\ssc P} (x)$-Hermitian $\frac{d^2}{dt^2} {x}^{(i)}$ given by 
$ \frac{1}{2m} \! \left(      \frac{d^2{x}^{i}}{dt^2}  +    \frac{d^2{x}^{\ssc I}}{dt^2}   \right)$.
We take the above as the quantum geodesic equations from the dynamical perspective.
Our result looks quite a bit more complicated than the vanishing of an expression like
$GE_a$ as given in Eq.(1), in the exact form as the classical result in terms of the
Christoffel symbol. However, it is still much simpler than the corresponding result from
DeWitt\cite{Dw}. Most importantly, there is no dependence on $\hbar$ or the particle mass
$m$. Simplification to the classical result, when all quantities commute with one another,
is certainly no problem. Yet, one can keep the more original form of the equations as in 
Eq.(\ref{dx}) and Eq.(\ref{dp}), or
  \bea&&
        \frac{d^2}{dt^2} {x}^{(i)} 
          = -  \frac{p_{\!\ssc H}}{m}  \frac{\partial_{\!\ssc J}  g^{{\ssc H}k}}{4} \frac{p_k}{m}  g^{{\ssc J}i} 
             +  \frac{p_{\!\ssc H}}{m}  \frac{\partial_{\!\ssc J}  g^{{\ssc H}i}}{4} g^{{\ssc J}k}  \frac{p_{k}}{m}  
             +  \frac{p_{\!\ssc K}}{m}  \frac{p_{\!\ssc H}}{m} g^{{\ssc H}j}       \frac{\partial_j g^{{\ssc K}i}}{4} + H.c.
  \eea
  
  We are now ready to look at the Hermiticity questions for the momentum components.
  There is a tension between those of $p^i$ and $p_i$, Hermiticity of one implies the
  lack of Hermiticity for the other. Hermiticity of $p^i$ is also generally tied to the Hermiticity
  of the velocity vector $\frac{dx^i}{dt}$ for a particle, as we generally expect the kinetic 
  term as proportional to the magnitude square of a vector that is the momentum $p_i$ 
  plus a term from a gauge potential that is a function only of the position observables 
  such as the (Hermitian) electromagnetic vector potential $A_i(x)$. From the 
  mathematical point of view, one can indeed introduce an involution, {\em i.e.} Hermitian 
  conjugation map, of the observable algebra as functions of the position and momentum 
  observables $x^i$ and $p^i$ satisfying the Heisenberg commutation relation or 
  Poisson bracket condition among $x^i$ and $p_i$ based on taking $x^i$ and $p^i$ 
  to themselves. That would be the  notion of the latter being Hermitian, we term 
  $g_{\!\ssc P}(x)$-Hermitian to emphasize the Hermiticity or Hermitian conjugation 
  map is based specifically on the choice of the canonical coordinates $x^i$ and $p_i$.
  and would not keep $p'^{a}$ Hermitian. And we certainly do not want a universal notion 
  of Hermiticity which would prefer basic observables under some frame of references
   in violation of the Relativity Principle.
  
  It is interesting to take a look back on the  question of whether one can have 
  a Schr\"odinger wavefunction representation fitting into our Heisenberg picture 
  framework. When one takes $p_i$ as the operator $-i\hbar \partial_i$, with the PMO 
  $g_{\!\ssc P}(x)$ as $\sqrt{|\det{g_{ij}(x)}|}$, 
  one has $-\hbar^2\Delta$ as $p_{\!\ssc I} g^{{\ssc I}j} p_j$. Yet, $p^i$ fails to be 
  $g_{\!\ssc P}(x)$-Hermitian. The Poisson bracket conditions say that $p_i$ must have 
  the form  $-i\hbar \partial_i f(x)=-i\hbar \partial_i -i\hbar (\partial_i f)$. No solution can 
  be obtained for the $g_{\!\ssc P}(x)$-Hermiticity of $p^i$. The wavefunction 
  $\phi(x) = \lla x | \phi \rra$ is mathematically an infinite set of complex coefficients
  expressing the state $\ket\phi$ as a linear combination of the position eigenstates
  $\ket{x}$.  However, even for the usual case with Euclidean $x^i$, we know that 
  the eigenstates are not truly vectors within the physical space of states on which
  one can have Hermitian position and momentum observables \cite{M}. Going to the
  general case of a curved geometry with an arbitrary coordinate system, we hardly
  know how to think about the position eigenstates and how one can justify the use
  of the corresponding Schr\"odinger wavefunction. So long as one does not stick to
  the idea that the space of the $x$ in $\phi(x)$ is the proper model of the physical
  space in the quantum theory, there is no good reason to assume the Schr\"odinger 
  wavefunction representation can be taken to a generic curved space(time).
  
 Some comments about a Lagrangian formulation or the variational problem in terms 
 of the invariant line element $\sqrt{dx^i g_{i{\ssc J}} dx^{\ssc J}}$ are in order. Firstly, 
 with all quantities involved being noncommutative ones, the line element should also
 be a noncommutative, q-number, quantity. Hence there is no direct application of the
 variational problem for the latter to obtain the dynamical picture we have, through the 
 Hamiltonian formulation. This is the case even for the case of simple quantum mechanics 
 in Euclidean space, for which the validity of the Heisenberg equation of motion should be 
 considered firmly established. 
 {The Hamiltonian evolution parameter in the equation of motion,} 
 as the parameter in the unitary time evolution is also the Hamiltonian evolution parameter 
 when seeing the Schr\"odinger equation as a set of Hamilton's equations for coordinates 
 of the (projective) vector space of states \cite{078}. A variational treatment taking the 
 free particle Lagrangian as $\frac{m}{2} \frac{dx^i}{dt}g_{i{\ssc J}} \frac{dx^{\ssc J}}{dt}$,
 however, cannot be performed without knowing the explicit action of the coordinate 
 derivatives of the kind {\boldmath $\partial_n$} for the position and velocity coordinate 
 observables. The same holds for formulating a Legendre transformation between 
 Lagrangian and Hamiltonian formulation. At this point, we have to leave the problem open.
 
 The free particle Hamiltonian, of course, can be extended to include terms from 
 interactions in the standard way as
 \bea
 H_t= \frac{1}{2m} (p_{\ssc I} -q A_{\ssc I}) g^{{\ssc I}i} (p_i -qA_i) + V(x) \;,
 \eea 
 where $A_I$ is a background electromagnetic vector potential and $V(x)$ a potential
 term. A system of particles with one Hamiltonian for the system having $V(x)$ involving 
 mutual interactions can be considered. Going to a full theory of electrodynamics of 
 charged particles is somewhat beyond quantum mechanics though.
 
 \section{Quantum Mechanics in Curved Spacetime}
  To think about quantum mechanics in curved spacetime, we can simply apply our
  analysis and results to the case of four-dimensional spacetime with a Minkowski 
  metric signature, hence a noncommutative symplectic geometry with four pairs of 
  canonical coordinate variables. The related issues about the picture of `relativistic'
  dynamics have been discussed in Refs.\cite{101,096}. We refrain from repeating 
  much of the details, only to highlight the following: For the classical case, the 
  mathematical geodesic problem for any manifold has the same form independent
  of the metric signature. So does the solution. The number of degrees of freedom 
  is exactly the dimension of the real manifold. After all a `relativistic' particle has
  a spacetime position with the four independent coordinates $x^\mu$. The 
  cotangent bundle of that spacetime manifold naturally gives a symplectic manifold 
  with a free momentum four-covector contributing the four canonical momentum 
  coordinates. It is the particle phase space that is Lorentz covariant at the `special 
  relativistic' limit, without any on-shell mass condition constraining the canonical 
  momentum four-vector. The free particle Hamiltonian dynamics in curved spacetime 
  given in the same form as the `nonrelativistic' three-dimensional case with a 
  Euclidean signature gives exactly the mathematical geodesic equations as the 
  solution. The Hamiltonian evolution parameter is  not the Newtonian time. Nor is 
  it the particle proper time, though for the free particle case {the latter can
  essentially be identified with it from the solution.} Note that the Hamiltonian itself 
  completely fixed the mathematical property of its evolution parameter anyway. 
  All results from any Hamiltonian in the more popular theory with the on-shell mass 
  condition can have an alternative formulation in the theory with the right matching
  Hamiltonian \cite{J}. Yet, more possibilities are allowed. There is no need to assume
  all particle proper times always agree with one another. The quantum mechanics we 
  talk about here is the quantum counterpart of such a classical theory. It has been 
  studied for a long time (see the books \cite{Hw,F}, and references therein). 
  The quantum theory is even more appealing than the classical one for a few reasons
   \cite{096}. 
   {Note that as in our explicit formulation here, the free particle 
     Hamiltonian as the magnitude square of the momentum four-vector is an invariant 
     quantity with the Hamiltonian equations of motion in fully covariant form. The
     quantum geodesic equations are independent of the particle mass. 
     The descriptions of a particle moving under a (constant) gravitational pull in flat
     spacetime and that of the geodesic motion in the accelerated frame match
     completely to one another when the equivalence of the gravitational mass
     and the inertial mass is taken. That is how one can get to the exact quantum 
     analog of the classical  consideration of the Weak Equivalence Principle \cite{101}. 
     With the  quantum position coordinates, at least the three spatial $x^i$, not being real
   number variables, the particle proper time certainly cannot be seen as one either, while 
   the Hamiltonian evolution parameter is always a real parameter.}   
         
 \section{Concluding Remarks}
 Flat space or spacetime is a vector space.  When physicists first introduced the 
 notion of a vector, it was about observables. In a curved manifold, there is no notion 
 of any position vector. Velocity and momentum vectors are then the basic vectors for 
 a particle. Particle velocity is naturally a tangent vector, at least for the classical case
 with spacetime as a real manifold. In the corresponding Hamiltonian formulation of 
 particle dynamics, the canonical momentum is naturally a covector. As such the
 canonical Poisson brackets and the Hamiltonian formulation from the symplectic
 geometry is metric independent. In classical physics and classical geometry, we
 have the notion of metric tensor as defining an inner product on the tangent space,
 and its inverse as giving the matching inner product on the cotangent space. Then
 we can have vector quantities with invariant magnitudes. The free particle Hamiltonian is 
 proportional to the magnitude square of the canonical momentum vector. If one 
cannot maintain all that in a quantum theory, then quantum physics should be 
considered so incompatible with our classical geometric notion of gravity and 
 spacetime that we certainly have to find the theory of quantum gravity with 
 a complete makeover of at least one of them. A key motivation of the study reported 
 here is to get to a picture of quantum mechanics in curved spacetime maintaining 
 a sensible notion of vector quantities as quantum observables, and the role of the 
 metric in defining the corresponding inner product. That is how we understand the 
 effect of spacetime curvature on particle dynamics. In our opinion, giving up all that 
 amounts to giving any physical notion of vector as well as giving up the basic picture 
 of gravity as given in GR. That is the main reason we are not satisfied with DeWitt's 
 formalism.
 
 Heisenberg and Dirac started the quantum theory with the idea of position and
 momentum observables as noncommutative, or q-number, canonical variables 
 instead of real number ones, giving the Poisson bracket as the commutator
 divided by $i\hbar$. Modern mathematics gives the quantum observable algebra 
 a noncommutative geometric picture, which may then be seen as the phase 
 space for the quantum system, as the exact analog of the classical case. From
 the perspective and beyond, there is no good reason to believe the classical 
 geometry behind the Schr\"odinger wavefunction representation can serve as 
 a valid model of our space, hence not necessary to commit to having such 
 a representation of quantum mechanics in any curved space or spacetime, or
 even simply for the no curvature case but in arbitrary system of curvilinear
 coordinates. Schr\"odinger's theory of wave mechanics was first accepted as
 an equivalent formulation of the Heisenberg-Dirac theory. Yet, that equivalence
 was only proven under the Cartesian coordinate picture, and even Schr\"odinger
 himself soon gave up the idea of wavefunction as a physical (matter) wave in
 space. Mathematically, it is simply a set of infinite numbers of coordinates for
 the state vector as a point in the Hilbert space. We set off to look at  the subject
 matter in terms of the observables, {\em i.e.} the physical quantities, along the
 original spirit of  Heisenberg and Dirac. The theory obtained is {\em not the same}
 as the one from DeWitt based on the Schr\"odinger representation, with what
 we argue to be more appealing theoretical features.
 
 Unlike the case of classical, commutative, geometry of real manifold, noncommutative 
 geometry as developed in mathematics hardly uses any notion of (local) coordinate 
 systems, not to say looking into coordinate transformations. 
 {For physicists, the local coordinate picture is important.} 
 Any explicit consideration of the Relativity Principle is about coordinate 
 transformations. Note that most of the notions of noncommutative geometry discussed in 
 the physics literature is, however, rather about plausible noncommutative space 
 with noncommutativity among position coordinates and not necessarily about quantum 
 theories. The key issues we have here have not been addressed there either. The
 practical case of quantum mechanics is the only established successful theory about 
 Nature that has naturally such a noncommutative geometry picture, though that has 
 not been considered much in the literature. Taking quantum mechanics to gravity, 
 or curved space(time), the geometric picture is especially important for the 
 application of the Relativity Principle, with the necessity to consider quantum 
 reference frame transformations. This article is however only  about
  reference frame transformations of the classical kind on the quantum position
  observables, within which we have already seen important nontrivial issues as
  addressed with strong implications about the class of noncommutative geometries 
  relevant to fundamental physics.  Extending our study to the quantum reference
  frame transformation is a big challenge to be taken up.
  
  With classical-like partial derivatives of quantum position observables introduced 
  as derivations through the Hamiltonian vector fields and classical-like prescriptions
  of transformation rules for vector and covector quantities, one needs further
  transformation rules for their Hermitian conjugate. Hermitian conjugate of a real 
  vector is its transpose. Though the components are the same,
   {the transposed vector is still 
  a different kind of vector with different properties under transformations.} 
  For a vector of non-Hermitian observables, the conjugate is completely different. 
  In curved space(time), the Hermitian conjugate or the transpose of a vector is not 
  a covector either. For vectors with components not commuting with those of the
  metric, one needs to pay special attention to that. With index notation to trace then 
  the four kinds of vectors, we have given a picture of quantum vector observables
 with invariant magnitude square in the same manner as in the classical theory
 including the role of the metric in that. A fully general notion of tensors may not be 
 possible though. The basic relation between a vector and its covector says that 
 one cannot have both having Hermitian components, the crucial tension is then 
 between the velocity vector $\frac{dx^i}{ds}$ ($s=t$ for the `nonrelativistic' case) 
 and momentum $p^i$ on one side and the canonical coordinate $p_i$ on the other. 
 We see having $p^i$ together with $x^i$ as $g_{\!\ssc P}(x)$-Hermitian, a notion 
 of Hermiticity that is reference frame or coordinate dependent, as the right choice. 
 No explicit PMO as $g_{\!\ssc P}(x)$ so long as the Heisenberg picture description 
 is concerned. 
 
 Generic (mixed) states for any algebra are defined as its functional. Pure states 
 as the extremal ones from the convex structure and be obtained \cite{AS}. With 
 a $C^*$-algebra, there is then a standard procedure to construct a Hilbert space 
  \cite{fD}. 
  {In the case that we have a Euclidean metric signature for our quantum
 observable algebra as formal functions of the position and momentum observables
 $x^i$ and $p^i$, with the involution taken as the notion of Hermitian conjugation
 based on our picture of the $g_{\!\ssc P}(x)$-Hermiticity,}
 all that should work along the line. For the case of a Minkowski metric signature, we 
 would have put in the extra Minkowski PMO $\eta$ we mentioned in the introduction,
 multiplying the $g_{\!\ssc P}(x)$. One may even argue that the usual  Schr\"odinger 
 picture may not be absolutely necessary. The pure state functionals are effectively
 expectation value functions of the states. They can be extended to give a picture
 of the states as evaluative homomorphisms of the observable algebra, each giving
 a noncommutative, q-number, value to each observable that mathematically 
 encodes the full information on the observable the quantum theory provides
  \cite{079,093}. The full statistical distribution of results from projective 
 measurement, for example, can be retrieved from that. Looking into the 
 Schr\"odinger picture and the explicit PMO, however, would provide more insight
 into the theoretical structure and may be practically convenient in applications. We
 certainly do not have the same inner product space for the Schr\"odinger states
 under different frames of reference. One may even doubt the validity of a universal
 vector space of states, or if that is rather a frame-dependent local description. 
 After all, in classical differential geometry, coordinate descriptions generally do
 not have global validity.
 
In conclusion, we have presented a new consistent formalism of quantum mechanics
in curved space(time)  with the metric $g_{ij}$ taken as quantum observables. The 
quantum geodesic equations given by either of the expressions of Eq.(\ref{geo}) are
independent of the mass of the (spin-zero) particle and obtained from the invariant
Hamiltonian $\frac{p_{\ssc I} g^{{\ssc I}j} p_j}{2m}$. The momentum observables
$p_i$ and $p_{\ssc I}$ are Hermitian conjugates of one another, both satisfying 
Heisenberg commutation relation with the position observables. Position-dependent
potential can be considered. While all position observables, or functions of them, are 
taken as generally Hermitian, the Hermiticity of the corresponding $p^i$ components 
has to be specific to the reference frame. Schr\"odinger representation is generally
not admissible. Some notions of tensors for quantum observables are established.  
The work suggests a new approach to quantum gravity completely giving up any 
notion of the classical, commutative, geometry of the position observables in the 
formulation. For example, the validity of our naive quantum Einstein equation 
$G_{\!a{\!\ssc A}} + \Lambda g_{a{\ssc\! A}} = 8 \pi G  T_{\!a{\ssc\! A}}$ should be 
looked into carefully.

 \section*{Acknowledgments}
The author is partially supported by research grants 
number  112-2112-M-008-019 and 113-2112-M-008-023 of the NSTC of Taiwan. 
He thanks P. Hajac and A. Sitarz for questions and comments.

\end{document}